\documentclass{article}
\usepackage{spconf,amsmath,epsfig}
\usepackage{bm}
\usepackage{booktabs}
\usepackage{float}
\usepackage{graphicx}
\usepackage{color}
\definecolor{brickred}{rgb}{0.8, 0.0, 0.0}
\usepackage{url}
\title{INTERPRETABLE ATTRIBUTED SCATTERING CENTER EXTRACTED VIA DEEP UNFOLDING}
\name{Haodong Yang\textsuperscript{1}, Zhe Zhang\textsuperscript{2,3}, Zhongling Huang\textsuperscript{1}\sthanks{*Corresponding Author.} \thanks{This work was supported by the National Natural Science Foundation of China under Grant 62101459 and the China Postdoctoral Science Foundation under Grant~BX2021248.}}
\address{
\textsuperscript{1}School of Automation, Northwestern Polytechnical University \\
\textsuperscript{2}Suzhou Aerospace Information Research Institute \\
\textsuperscript{3}Aerospace Information Research Institute, Chinese Academy of Sciences (CAS)}

\begin{document}
%\ninept
%
\maketitle
\begin{abstract}

Most existing sparse representation-based approaches for attributed scattering center (ASC) extraction adopt traditional iterative optimization algorithms, which suffer from lengthy computation times and limited precision. This paper presents a solution by introducing an interpretable network that can effectively and rapidly extract ASC via deep unfolding. Initially, we create a dictionary containing reliable prior knowledge and apply it to the iterative shrinkage-thresholding algorithm (ISTA). Then, we unfold ISTA into a neural network, employing it to autonomously and precisely optimize the hyperparameters. The interpretability of physics is retained by applying a dictionary with physical meaning. The experiments are conducted on multiple test sets with diverse data distributions and demonstrate the superior performance and generalizability of our method.

\end{abstract}
\begin{keywords}
Attributed Scattering Center, Deep Unfolding, Sparse Representation
\end{keywords}
\section{Introduction}
\label{sec:intro}

Synthetic Aperture Radar (SAR) is an active microwave sensor capable of capturing images in all-day all-weather conditions. Since the capability of the ASC model to parameterize SAR target characteristics, there has been an increasing amount of research focused on extracting ASC \cite{wen2019new, li2014sparse} and employing it for interpreting SAR data \cite{huang2024physics, WC2024}.

\begin{figure}
    \centering
    \includegraphics[width=7.5cm]{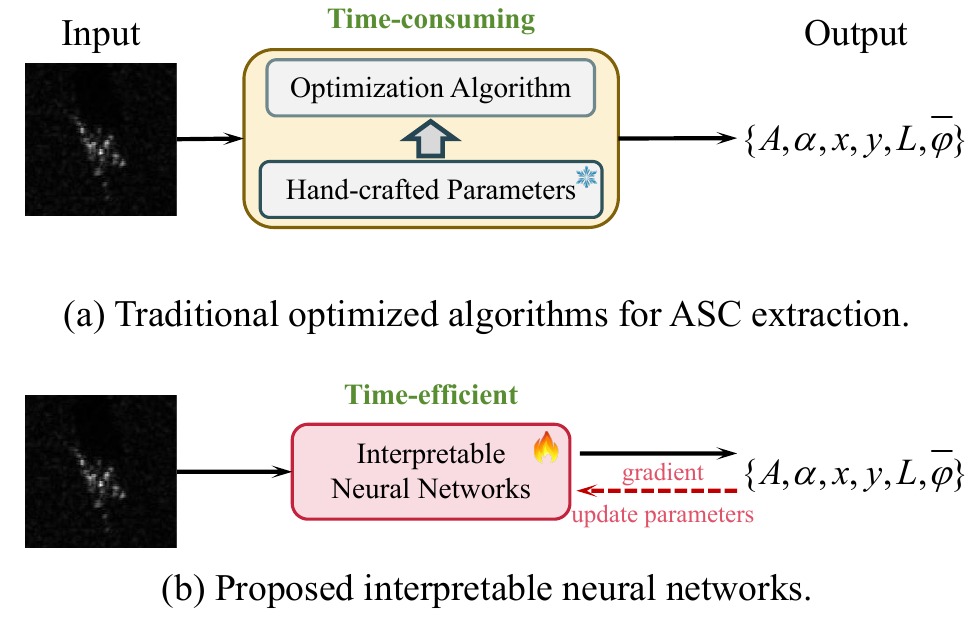}
    \caption{Compared with the traditional optimized algorithms and the proposed method for ASC extraction.}
    \label{fig:motivation}
\end{figure}

The ASC parameter estimation process can be considered as a sparse representation problem due to the sparsity of radar echo. Several studies \cite{wen2019new, li2014sparse} adopt sparse optimization methods to extract ASC, with the primary algorithms employed being orthogonal matching pursuit (OMP) \cite{OMP}, approximate message passing (AMP) \cite{AMP} and ISTA \cite{ISTA}. As shown in Fig. \ref{fig:motivation} (a), traditional approaches frequently execute iterative optimization on single input data and require manual adjusting of hyperparameters. This will encounter issues such as inefficiency and imprecise results. 

To tackle the aforementioned problems, we unfold ISTA into an interpretable neural network to achieve efficient and credible extraction of ASC, as shown in Fig. \ref{fig:motivation} (b). The main contributions are as follows:

\begin{enumerate}

  \item We propose a novel approach via deep unfolding to extract ASC efficiently and precisely. In contrast to current deep unfolding methods employing random dictionaries \cite{zhang2023physics}, we utilize a dictionary that possesses physics prior information to enhance the model's interpretability.
  
  \item Extensive experiments are conducted on the Moving and Stationary Target Acquisition and Recognition (MSTAR) dataset \cite{MSTAR} to demonstrate the efficacy of the proposed method. Concretely, we train and validate the model using data at a depression angle of 17°, and then generalize it within data at depression angles of 15° and 45°. The results demonstrate the proposed approach has superior generalization. 
  
\end{enumerate}

% This work's sections follow this structure. The ASC model is briefly introduced in Section \ref{sec:ASCM}. Sections \ref{sec:mtd} and \ref{sec:exp} introduce the proposed and conducted experiments, along with commentary. The conclusion is in Section \ref{sec:concls}.

\section{ATTRIBUTED SCATTERING CENTER MODEL}
\label{sec:ASCM}
%%%% 写ASC模型基础

The ASC model is presented based on geometric theory diffraction and physical optics theory \cite{ASC} to analyze the electromagnetic scattering response of the target. It can be viewed as a linear combination of $N$ individual ASCs, each of which is regarded as a function of radar frequency $f$ and aspect angle $\varphi$, where $f\in (f_c - B/2, f_c + B/2)$ and $\varphi \in (-\varphi_{syn}/2, \varphi_{syn}/2)$ with $B$, $f_c$ and $\varphi_{syn}$ denoting the bandwidth, the center frequency and the synthetic angle respectively. The ASC model at small angles can be modeled as follows \cite{mengdao2022electromagnetic}: 
%% 在小角度下可以被建模成以下\cite{}
\begin{equation}
\label{equ:ASC}
\begin{split}
    \bm{E(f,\varphi;\Theta)} = & \sum\limits_{i=1}^{K_0} \bm{E_i(f,\varphi;\theta_i)} \\
    = &\sum\limits_{i=1}^{K_0} A_i \cdot (j\frac{\bm{f}}{f_c})^{\alpha_i} \\
    & \cdot \operatorname {exp}(-j\frac{4\pi\bm{f}}{c}(x_i \cos{\bm{\varphi}} + y_i \sin{\bm{\varphi}}) \\
\end{split}
\end{equation}
%% E提上去
\noindent where $\bm{E}$ is the radar echo. $\bm{\Theta} = [\bm{\theta}_1, \bm{\theta}_2,\cdots,\bm{\theta}_{K_0}]$ denotes the ASC parameter set and $K_0$ is the number of ASCs. $\bm{E_{i}}$ represents the $i$th ASC. $j = \sqrt{-1}$ indicates the imaginary unit and $c$ is the propagation velocity. $ \bm{\theta_i}=\{A_i,\alpha_i,x_i,y_i\}$ is the $i$th ASC parameter set. $A_i$ indicates the complex amplitude. $x_i$ and $y_i$ represent the coordinates in the range and azimuth directions, respectively. $\alpha$ reflects the frequency dependency factor that can distinguish the geometric structure. 

Given the sparsity of radar echo in the ASC parameter space, the theory of sparse signal representation enables the analysis and extraction of ASC. Thus, the ASC model in Eq. \ref{equ:ASC} can be explained as below:
\begin{equation}
\label{equ:Sp}
\begin{split}
    \widetilde{s} = \bm{\widetilde{\Phi}(x,y)}z
\end{split}
\end{equation}
where $\widetilde{s}$ represents the vectorization of $\bm{E}$ and $z$ is the sparse coefficient vector representing $ A(j\frac{\bm{f}}{f_c})^{\alpha} $ in Eq. \ref{equ:ASC}. $\bm{\widetilde{\Phi}}$ indicates a parameterized dictionary that includes the position information of ASCs in the signal domain, which corresponds to the exponent term in Eq. \ref{equ:ASC}. 
$\bm{f}$ and $\bm{\varphi}$ can be considered as vectors that are uniformly sampled $P$ and $Q$ times within their corresponding ranges, respectively. $\bm{x}$ and $\bm{y}$ are consistently sampled $M$ and $N$ times respectively. Thus, the dimensions of $\bm{\widetilde{\Phi}}$ are $PQ \times MN$. The dictionary can be formulated in columns as follows:
%% 在这里引入f和phi的采样
%% s改一下
\begin{equation}
\label{equ:D}
\begin{aligned}
    &\bm{\widetilde{\Phi}(x,y)} = {\widetilde{\Phi}_{:,1},\widetilde{\Phi}_{:,2},\cdots,\widetilde{\Phi}_{:,MN}},\\
    &\widetilde{\Phi}_{:,MN} = vec(\operatorname {exp}(-j\frac{4\pi\bm{f}}{c}(x_M \cos{\bm{\varphi}} + y_N \sin{\bm{\varphi}})).
\end{aligned}
\end{equation}

\begin{figure}[!ht]
    \centering
    \includegraphics[width=7.5cm]{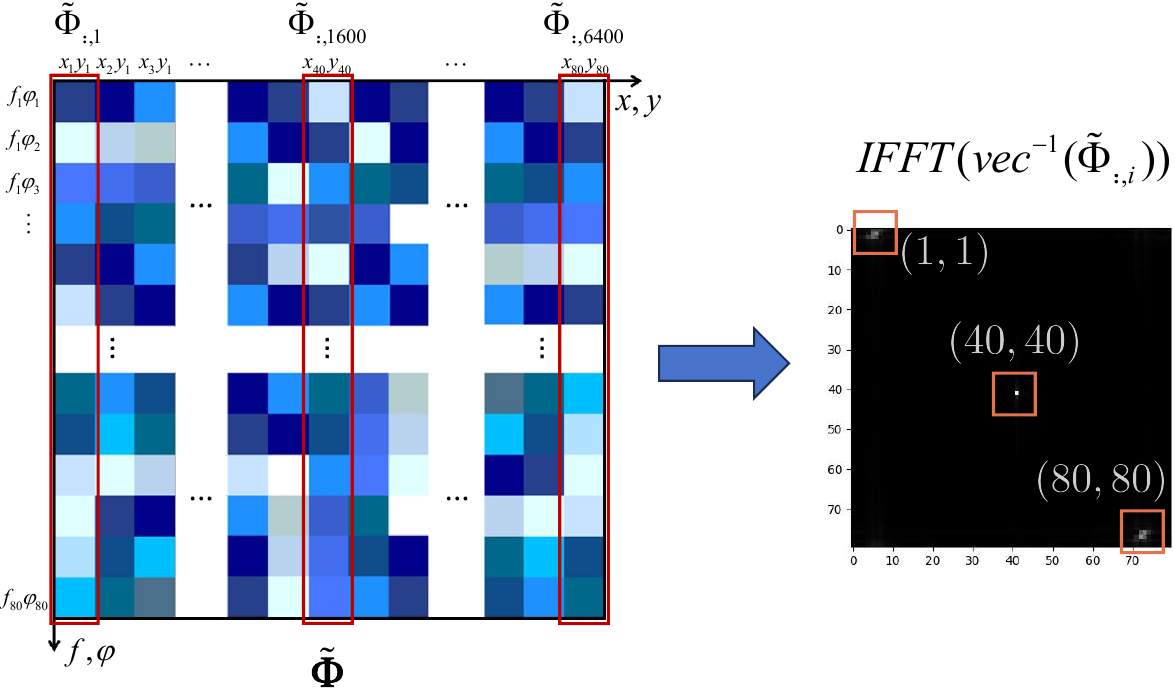}
    \caption{The visualization of each column in the dictionary within the image domain.}
    \label{fig:dictionary}
\end{figure}

Considering the targets in MSTAR fall inside the central $80 \times 80$, we construct the dictionary specifically for this region, in which the sampling numbers P, Q, M, and N are all set to 80. Fig. \ref{fig:dictionary} illustrates the significance of each column in the dictionary within the image domain. 

Each column of the dictionary in the image domain can be represented as the scattering center of the corresponding coordinates via the Inverse Fast Fourier Transform (IFFT) and inverse vectorization, as illustrated in Fig. \ref{fig:dictionary}. Then the dictionary in the image domain can be approximated as a sparse matrix where the echo of each column is concentrated on the diagonal. Thus, extracting ASC in the image domain is essentially a matching problem. 

Based on sparse representation theory, $z$ can be solved in the image domain by the following step:
\begin{equation}
\label{equ:image domain}
    \mathop{\arg\min}\limits_{z}||\bm{\Phi}z-s||_2+\lambda||z||_1, 
\end{equation}
where $\bm{\Phi}$ and $s$ are the dictionary in the image domain and the complex-valued image. $\lambda$ is a hyperparameter. In this paper, we adopt ISTA to optimize it.

\section{Methods}
\label{sec:mtd}
\begin{figure*}[!ht]
    \centering
    \includegraphics[width=\linewidth]{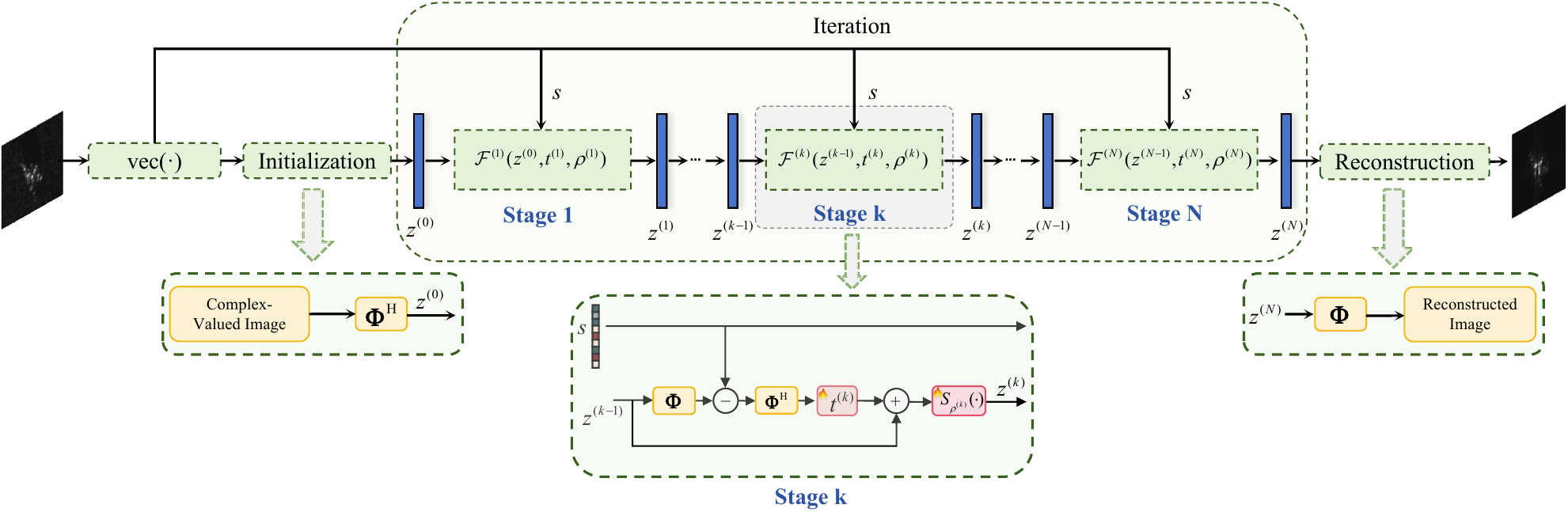}
    \caption{The overall architecture of the proposed method.}
    \label{fig:overview}
\end{figure*}

%%% 写传统ISTA算法和本文展开的方法
In this section, we initially provide a concise overview of ISTA. Then we unfold it into a neural network and explain its operation. The architecture is shown in Fig. \ref{fig:overview}.

\subsection{ISTA}

ISTA is a widely utilized approach for proximal gradient descent that has been implemented in numerous problems involving sparse representation. It can be applied to solve the optimization problem presented in Eq. \ref{equ:image domain} by following the steps below:

\begin{equation}
\label{equ:tradISTA}
\begin{split}
    & x^{(k)} = z^{(k-1)} - t\bm{\Phi^}{H}(\bm{\Phi} z^{(k-1)}-s), \\
    & z^{(k)} = S_\rho(x^{(k)}).
\end{split}
\end{equation}
\noindent The variable $k$ represents the index of the iteration, whereas $t$ represents the step size. The soft-thresholding function for complex valued data, denoted as $S_\rho$, can be defined as follows:

\begin{equation}
\label{equ:tradISTA}
\begin{split}
    & S_\rho(x) = sign(x)max(|x|-\rho,0), \\
    & sign(x) = 
    \left\{ 
    \begin{aligned}
    &\frac{x}{|x|}, &|x|>0, \\
    &0, &|x|=0,
    \end{aligned}
    \right.
\end{split}
\end{equation}
\noindent where $\rho$ is a hyperparameter. The method requires handcrafted parameters $t$ and $\rho$ based on the input $s$, resulting in a time-consuming and labor-intensive process. 

\subsection{Interpretable Unfolding ISTA}

To perform efficient ASC parameter extraction, we unfold ISTA into a neural network and automatically optimize the hyperparameters. Different from existing deep unfolding methods \cite{zhang2018ista, zhang2020amp} that employ random matrices like Gaussian matrix as the dictionary, we adopt $\bm{\Phi}$ as it encompasses physical information of radar echo. The proposed approach can be categorized into three phases: Initialization, Iteration, and Reconstruction, as depicted in Fig. \ref{fig:overview}.

\textbf{Initialization.}
The complex-valued image is primarily vectorized before the initialization phase, denoted as $s$. The initial value $z^{(0)}$ of Stage 1 can be calculated as $z^{(0)} = \bm{\Phi}^H s$.

\textbf{Iteration.}
The iterative component consists of N analogous phases, with each phase being considered as an amplified version of ISTA. The internal intricacies of each phase are identical to those of Stage K depicted in the Fig. \ref{fig:overview}, which can be formulated as follows:
\begin{equation}
\label{equ:unfoldISTA}
z^{(k)} = S_{\rho^{(k)}}(z^{(k-1)} + t^{(k)}\bm{\Phi}^{H}(\bm{\Phi} z^{(k-1)}-s)).
\end{equation}

In contrast to traditional ISTA, we introduce $t$ and $\rho$ as trainable parameters in every stage. The quantity of parameters requiring updates and the inference time are contingent upon the overall stages $N$. We set $N$ to be 4 empirically, resulting in $8$ parameters total. The hyperparameters $t$ and $\rho$ for each stage are set to $0.01$ and $0.005$ respectively.

\textbf{Reconstruction.}
The reconstructed image can be obtained by $z^{(N)}$ through the following procedures:
\begin{equation}
\label{equ:recon}
\hat{s} = \bm{\Phi} z^{(N)},
\end{equation}
where $\hat{s}$ denotes the reconstructed image. If $z^{(N)}$ is sufficiently sparse, then $\hat{s}$ can be regarded as the outcome of eliminating the noisy regions of $s$ while preserving the area with prominent scattering characteristics.

Following other deep unfolding methods \cite{zhang2018ista, zhang2020amp}, we define the loss as follows for each sample:
\begin{equation}
\label{equ:loss}
Loss = ||s-\hat{s}||_2+\lambda||z^{(N)}||_1.
\end{equation}
$||s-\hat{s}||_2$ indicates the residual loss, which is calculated by L2 normalization. $\lambda||z^{(N)}||_1$ serves as a regularization component. $\lambda$ is a hyperparameter and is set to be 300 empirically.

\section{Experiments}
\label{sec:exp}
%% 1、介绍数据集和实验设置
%% 2、展开方法与传统方法的对比
%% 3、lambda参数的消融实验
%% 4、
\subsection{Dataset and Experimental Settings}
The experiments are conducted on the MSTAR dataset \cite{MSTAR}. We randomly select 10\% of the 17° depression angle samples as the training set and select another 10\% of the remaining samples as the validation set. The test data for the 15° depression angle and the 45° depression angle are called D-15 and D-45 respectively.
% with 2747 slices of 17° depression angle and 2456 slices of 15° depression angle.

We employ AdamW \cite{AdamW} along with the OneCycleLR \cite{LR} to optimize the model. The weight decay is $0.05$ and the learning rate is $2e-3$. The epoch and the batch size are $50$ and $16$ respectively. The studies are performed on a GeForce RTX 3090. The slices are normalized by the L2 normalization and center cropped to $80\times80$. 

\subsection{Ablation Study}

\begin{table}[!ht]\tiny
     \caption{The residual loss and inference time with various $N$ on D-15.}
        \label{tab:s_n}
    \resizebox{\linewidth}{!}
    {
        \begin{tabular}{c c c c c}
        \toprule
         Stage Number & 2 & 4 & 6 & 8 \\
        \midrule
        Residual Loss  &0.6753 & 0.6617 & 0.6601 & 0.6603\\
        Inference Time (s)  & 0.0616 & 0.0729 & 0.0887 & 0.0942\\
        \bottomrule
        \end{tabular}}
  \end{table}

The residual loss and inference time on D-15 are documented in Table \ref{tab:s_n} as the number of network stages $N$ increases incrementally. Once the number of stages surpasses $4$, the decrease in residual loss becomes negligible, while the model inference time gradually grows. In summary, we decided to set $N$ to $4$ in the following experiments.

\begin{figure}[!ht]
    \centering
    \includegraphics[width=\linewidth]{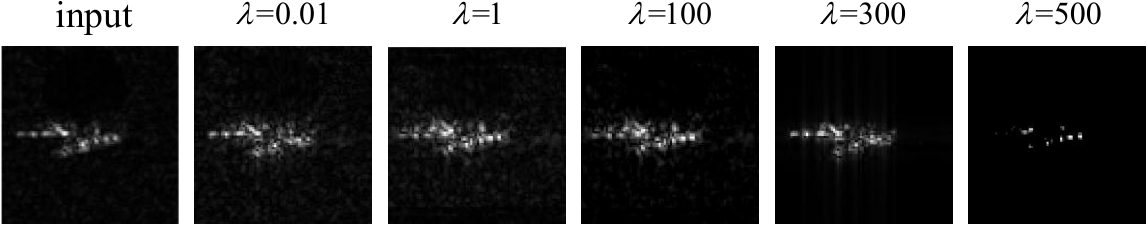}
    \caption{The reconstruction image within various $\lambda$.}
    \label{fig:lambda}
\end{figure}

The images reconstructed within various $\lambda$ are displayed in Fig. \ref{fig:lambda}. As $\lambda$ grows to $300$, the background noise interference steadily reduces, resulting in an improvement in the quality of the reconstructed image. However, when $\lambda$ is $500$, it is over optimized. Therefore, $\lambda$ is set to $300$ in the following experiments.

\subsection{Comparison with Traditional Methods}

Table \ref{tab:recon} compares the proposed and traditional approaches on D-15, displaying average residual loss and inference time for each image. The sparsity of the OMP algorithm is 40. ISTA's $t$ and $\rho$ align with the unfolding network at $0.01$ and $0.005$, respectively. The change rate between adjacent iterations of the AMP is set to $0.1$.

\begin{table}[!ht]\tiny
     \caption{The test results of various methods on D-15. The best are in \textbf{bold}.}
        \label{tab:recon}
    \resizebox{\linewidth}{!}
    {
        \begin{tabular}{c c c c c}
        \toprule
         Method & AMP & OMP & ISTA & Our method \\
        \midrule
        Residual Loss  & 1.4993 & 1.1436 & 0.9440 & \textbf{0.6617}\\
        Inference Time (s)  & 84.5331 & 199.1839 & 73.6324 & \textbf{0.0729}\\
        \bottomrule
        \end{tabular}}
  \end{table}
  
Our method demonstrates a significant reduction in residual loss compared to traditional methods, ranging from $29.9\%$ (compared to ISTA) to $55.9\%$ (compared to OMP). In addition, our method can speed up the inference time by nearly 100 times ($73.6324\rightarrow0.0729$). Collectively, the proposed approach can extract ASC rapidly and precisely.

\begin{figure}[!ht]
    \centering
    \includegraphics[width=7.5cm]{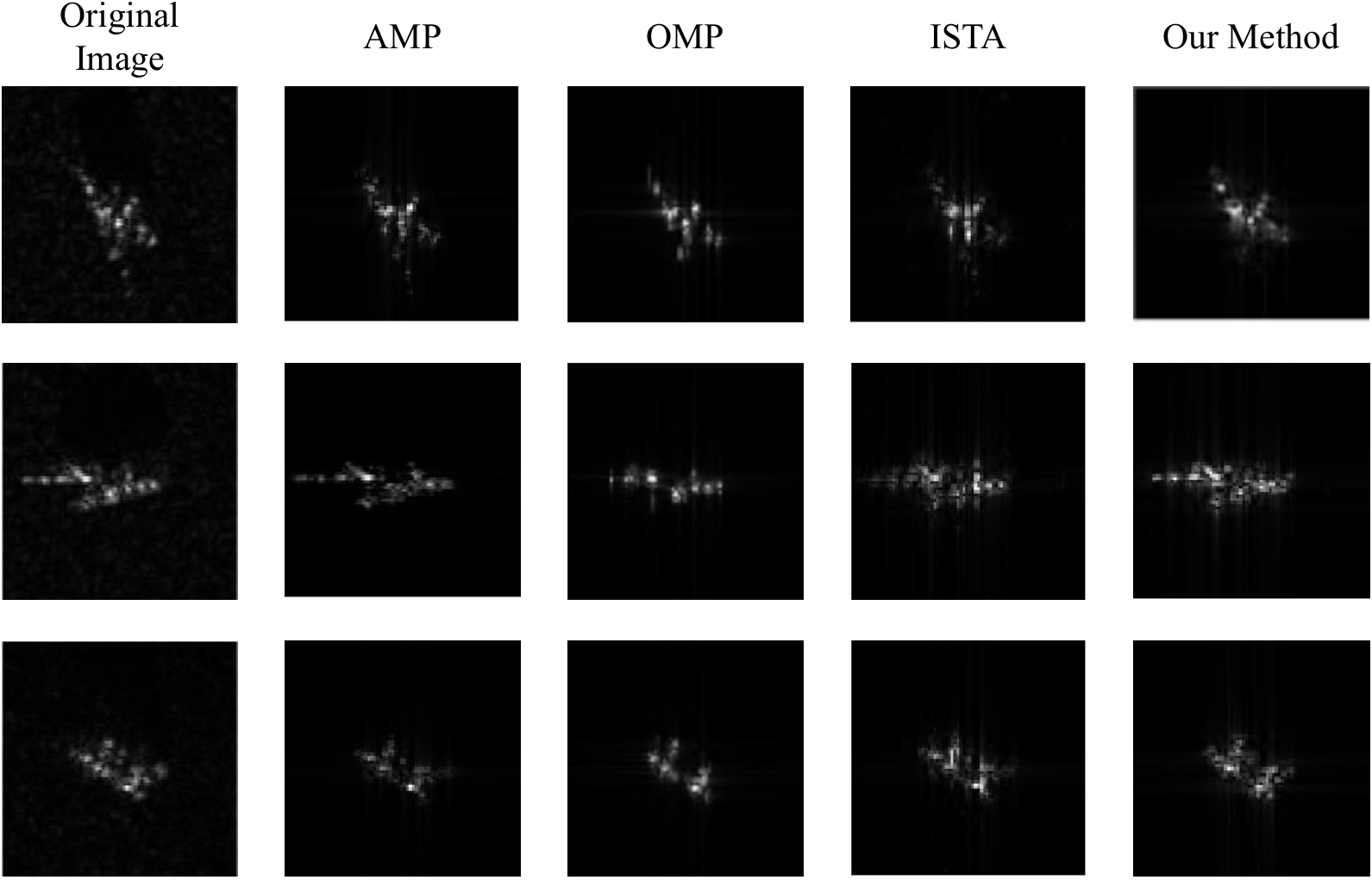}
    \caption{Comparison of the reconstructed images between our method and traditional methods.}
    \label{fig:recon}
\end{figure}

Fig. \ref{fig:recon} presents the reconstructed images comparison between our method and traditional methods. It is evident that the proposed method outperforms the traditional algorithm in terms of accurately identifying and reconstructing the target area. 

\subsection{Model Generalization Discussion}

\begin{figure}[H]
    \centering
    \includegraphics[width=\linewidth]{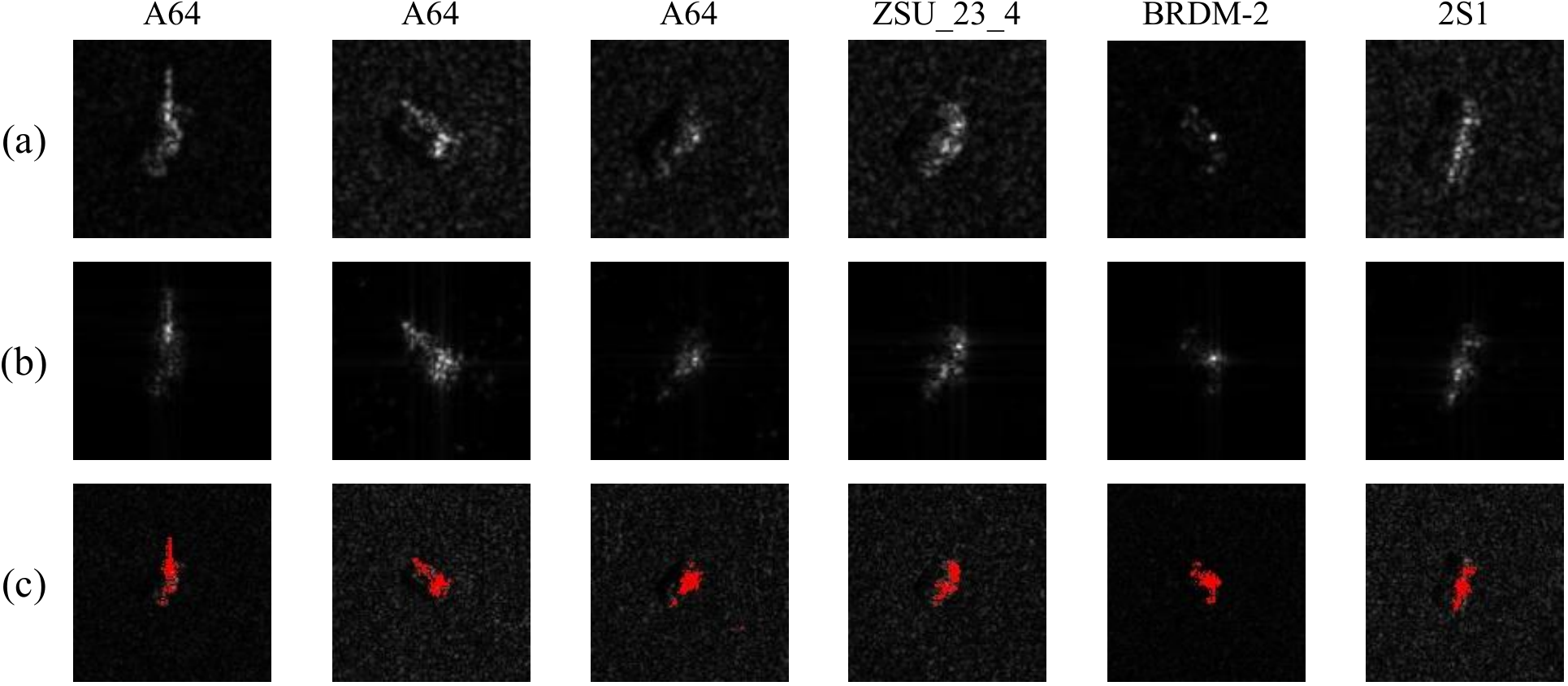}
    \caption{The test results on D-45. (a)-(c) represents the original complex-valued image, the reconstructed image and the ASC visualization respectively. }
    \label{fig:general}
\end{figure}

Fig. \ref{fig:general} displays the ASC extraction results on D-45. It is worth noting that category A64 is not in the training set, as shown by the first three columns in the figure. The findings indicate that when the network is trained on data with a depression angle of 17°, it demonstrates impressive performance on the test set with a depression angle of 45°. Both the reconstruction image and the ASC visualization emphasize that the areas contain significant scattering characteristics, which demonstrates the exceptional ability of the model's generalization.

\section{Conclusion}
\label{sec:concls}

Existing sparse representation theory-based ASC extraction methods adopt iterative optimization algorithms, which are slow and imprecise. We propose an interpretable deep unfolding method to achieve ASC extraction rapidly and precisely. Empirical evidence indicates that the proposed strategy can surpass traditional methods among test sets with diverse data distributions. Our future study will prioritize extracting the last parameters and acquiring all ASC parameter values. 
% References should be produced using the bibtex program from suitable
% BiBTeX files (here: strings, refs, manuals). The IEEEbib.bst bibliography
% style file from IEEE produces unsorted bibliography list.
% -------------------------------------------------------------------------
\bibliographystyle{IEEEbib}
\bibliography{strings,refs}

\begin{thebibliography}{10}

\bibitem{wen2019new}
J.~Wen and Z.~Wang,
\newblock ``A new method for parameter estimation of attributed scattering centers based on amplitude-phase separation,''
\newblock {\em Journal of Radars}, vol. 8, no. 5, pp. 606--615, 2019.

\bibitem{li2014sparse}
F.~Li, B.~Jiu, H.~Liu, Y.~Wang, and L.~Zhang,
\newblock ``Sparse representation based algorithm for estimation of attributed scattering center parameter on sar imagery,''
\newblock {\em Journal of Electronics \& Information Technology}, vol. 36, no. 4, pp. 931--937, 2014.

\bibitem{huang2024physics}
Z.~Huang, C.~Wu, X.~Yao, Z.~Zhao, X.~Huang, and J.~Han,
\newblock ``Physics inspired hybrid attention for sar target recognition,''
\newblock {\em ISPRS Journal of Photogrammetry and Remote Sensing}, vol. 207, pp. 164--174, 2024.

\bibitem{WC2024}
Z.~Huang, C.~Wu, X.~Yao, L.~Wang, and J.~Han,
\newblock ``Physically explainable intelligent perception and application of sar target characteristics based on time-frequency analysis,''
\newblock {\em Journal of Radars}, vol. 12, pp. 1--14, 2023.

\bibitem{OMP}
J.~A. Tropp and A.~C. Gilbert,
\newblock ``Signal recovery from random measurements via orthogonal matching pursuit,''
\newblock {\em IEEE Transactions on information theory}, vol. 53, no. 12, pp. 4655--4666, 2007.

\bibitem{AMP}
D.~L. Donoho, A.~Maleki, and A.~Montanari,
\newblock ``Message-passing algorithms for compressed sensing,''
\newblock {\em Proceedings of the National Academy of Sciences}, vol. 106, no. 45, pp. 18914--18919, 2009.

\bibitem{ISTA}
T.~Blumensath and M.~E. Davies,
\newblock ``Iterative thresholding for sparse approximations,''
\newblock {\em Journal of Fourier analysis and Applications}, vol. 14, pp. 629--654, 2008.

\bibitem{zhang2023physics}
J.~Zhang, B.~Chen, R.~Xiong, and Y.~Zhang,
\newblock ``Physics-inspired compressive sensing: Beyond deep unrolling,''
\newblock {\em IEEE Signal Processing Magazine}, vol. 40, no. 1, pp. 58--72, 2023.

\bibitem{MSTAR}
{Sandia National Laboratory},
\newblock ``{The Air Force Moving and Stationary Target Recognition Database},'' \url{https://www.sdms.afrl.af.mil/index.php?collection=mstar}, {Accessed May 2023}.

\bibitem{ASC}
M.~J. Gerry, L.~C. Potter, I.~J. Gupta, and A.~Van~Der Merwe,
\newblock ``A parametric model for synthetic aperture radar measurements,''
\newblock {\em IEEE transactions on antennas and propagation}, vol. 47, no. 7, pp. 1179--1188, 1999.

\bibitem{mengdao2022electromagnetic}
M.~Xing, Y.~Xie, Y.~Gao, J.~Zhang, J.~Liu, and Z.~Wu,
\newblock ``Electromagnetic scattering characteristic extraction and imaging recognition algorithm: A review,''
\newblock {\em Journal of Radars}, vol. 11, no. 6, pp. 921--942, 2022.

\bibitem{zhang2018ista}
J.~Zhang and B.~Ghanem,
\newblock ``Ista-net: Interpretable optimization-inspired deep network for image compressive sensing,''
\newblock in {\em Proceedings of the IEEE conference on computer vision and pattern recognition}, 2018, pp. 1828--1837.

\bibitem{zhang2020amp}
Z.~Zhang, Y.~Liu, J.~Liu, F.~Wen, and C.~Zhu,
\newblock ``Amp-net: Denoising-based deep unfolding for compressive image sensing,''
\newblock {\em IEEE Transactions on Image Processing}, vol. 30, pp. 1487--1500, 2020.

\bibitem{AdamW}
I.~Loshchilov and F.~Hutter,
\newblock ``Decoupled weight decay regularization,''
\newblock {\em arXiv preprint arXiv:1711.05101}, 2019.

\bibitem{LR}
L.~N. Smith and N.~Topin,
\newblock ``Super-convergence: Very fast training of neural networks using large learning rates,''
\newblock vol. 11006, pp. 369--386, 2019.

\end{thebibliography}

\end{document}